\documentclass[12pt]{article}

\usepackage{moreverb}

\usepackage{csquotes}
\usepackage[lofdepth,lotdepth]{subfig}

\usepackage{amssymb}
\usepackage{amsthm}
\usepackage{amsmath}
\usepackage{graphicx}
\usepackage{cite}
\usepackage{caption}

\usepackage{booktabs}
\usepackage{threeparttable}
\usepackage{longtable}
\usepackage{subfloat}
\usepackage{nth}
\usepackage{multirow}
\usepackage{times}
\usepackage{bm}
\usepackage{verbatim}
\usepackage{tabularx}
\newcolumntype{Y}{>{\centering\arraybackslash}X}
\usepackage{breqn}

\usepackage{color}
\newcommand{\indep}{\rotatebox[origin=c]{90}{$\models$}}

\theoremstyle{plain}
\newtheorem{theorem}{Theorem}
\newtheorem{prop}[theorem]{Proposition}

\begin{document}

\title{Robust Estimation of the Weighted Average Treatment Effect for A Target Population}
\author{Yebin Tao and Haoda Fu \\
	\emph{Lilly Corporate Center} \\
	\emph{Eli Lilly and Company, Indiana 46285, USA} \\
	\textsl{tao\_yebin@lilly.com; fu\_haoda@lilly.com}
	}

\date{\today}
\maketitle

\begin{abstract}
The weighted average treatment effect (WATE) is a causal measure for the comparison of interventions in a specific target population, which may be different from the population where data are sampled from. For instance, when the goal is to introduce a new treatment to a target population, the question is what efficacy (or effectiveness) can be gained by switching patients from a standard of care (control) to this new treatment, for which the average treatment effect for the control (ATC) estimand can be applied. In this paper, we propose two estimators based on augmented inverse probability weighting to estimate the WATE for a well defined target population (i.e., there exists a target function that describes the population of interest), using observational data. The first proposed estimator is doubly robust if the target function is known or can be correctly specified. The second proposed estimator is doubly robust if the target function has a linear dependence on the propensity score, which can be used to estimate the average treatment effect for the treated (ATT) and ATC. We demonstrate the properties of the proposed estimators through theoretical proof and simulation studies. We also apply our proposed methods in a comparison of glucagon-like peptide-1 receptor agonists therapy and insulin therapy among patients with type 2 diabetes, using the UK clinical practice research datalink data.
\end{abstract}


\section{Introduction}

In causal inference, the average treatment effect (ATE) is a standard measure for the comparison of interventions. It is defined on the entire population where data are sampled from. In certain scenarios, the target population may be either a subpopulation or a population with different covariate distributions. Our motivating example is an
observational study to compare the effectiveness of glucagon-like peptide-1 receptor agonists (GLP-1 RA) therapy (treatment) and insulin therapy (control) for patients with type 2 diabetes (T2D). Due to the weight loss benefit of GLP-1 RA therapy \cite{wang2011}, physicians tend to prescribe it more often for patients with higher body mass index
(BMI) as their first-line injectable anti-diabetic treatment. To address questions on the effectiveness of using GLP-1 RA therapy if we switch first-line injectable insulin users to GLP-1 RA therapy, a useful estimand is the ATE for the control (ATC). Generally, when the goal is to introduce a new treatment to a target population, the question is what efficacy (or effectiveness) can be gained by switching patients from a standard of care to this new treatment, in which case, the ATC estimand should be considered. However, when the primary interest is in the safety problems caused by a certain medication, the question is about the difference in adverse events if patients who currently take the medication switch to a standard of care, in which case, the ATE for the treated (ATT) estimand should be applied. Furthermore, there is the issue of external validity, i.e., the generalizability of a random sample to a target population \cite{stuart2011, stuart2015, pearl2014}. Due to selection bias or other difficulties in collecting samples, the study population may be different from the prespecified target population. Therefore, adjustments have to be made for valid causal inference about the target population. Assuming sufficient overlap in key features between the sampled data and the target population, we can weight the samples according to the covariate distributions of the target population in order to make valid causal inference, leading to the weighted ATE (WATE) \cite{hirano2003}.

Various balancing weights have been proposed for the estimation of the WATE in a prespecified target population, in addition to the ATT and ATC. A useful example is the overlap weights, with a patient's probability of being sampled proportional to the product of the treatment probability and the control probability \cite{vansteelandt2014, li2014}. Patients who, based on their characteristics, are equally likely to be administered treatment or control in standard practice are given a larger representation in the study. The estimand by overlap weighting is called the ATE for the overlap population (ATO), which, of all WATEs, achieves the smallest asymptotic variance as well as exact balance of covariates between two treatment groups \cite{crump2006, li2014}. 

In observational studies, certain assumptions are required for the identification of the WATE. In order to make internally valid causal interpretation, one has to adjust for confounding, which results from potential covariate imbalance between the two treatment arms. Commonly used methods to adjust for covariate imbalance include regression, matching (e.g., \cite{rosenbaum1985} and \cite{rosenbaum1989}) and weighting (e.g., \cite{hirano2001} and \cite{hirano2003}), with the last two relying on the propensity score, defined as the conditional probability of receiving treatment given covariates. Due to concerns about model misspecification in those methods, recently there has been substantive interest in doubly robust estimators, which involve models for both the propensity score and the conditional mean of the outcome, and remain consistent if either model is correctly specified. The original doubly robust estimator is constructed by augmented inverse probability weighting (AIPW) in the missing data setting \cite{robins1994missing}. Later, a number of doubly robust estimators have been proposed in both missing data and causal inference settings (e.g., \cite{bang2005}, \cite{cao2009}, \cite{tan2010} and \cite{rotnitzky2012}). However, the majority of these doubly robust estimators in the causal inference setting are for the ATE. To our knowledge, very few studies have worked on doubly robust estimators for the WATE (e.g., Kaiser \cite{kaiser2016} on the ATT for nonnegative outcome).

In this paper, we propose two estimators based on AIPW for the WATE that enjoy double robustness (i.e., consistent if either the propensity score or the conditional mean model is correctly specified). Given a known or correctly specified function that indicates the target population, the first proposed estimator is doubly robust. It is also more efficient than the estimator based on inverse probability weighting (IPW) if both the propensity score and the conditional mean models are correctly specified. When the function for the target population is subject to misspecification, for example, if it depends on the propensity score, the first estimator is no longer doubly robust. However, in the case of the target function linearly dependent on the propensity score, the second proposed estimator is doubly robust, which can be used to estimate the ATT and ATC. 

The remainder of the paper is organized as follows. In Section 2, we introduce notation and assumptions to identify the WATE, and propose robust estimators. We then demonstrate in Section 3 the properties of the proposed estimators through simulation studies. In Section 4, we apply the proposed estimators to the motivating example to compare the GLP-1 RA therapy and insulin therapy for T2D patients, using the UK clinical practice research datalink (CPRD) data. Finally, we summarize our main contributions and discuss about future development in Section 5.

\section{Weighted average treatment effect (WATE)}

\subsection{Definition and identifiability}
Consider a random sample of size $n$ from a population of interest. Let $A$ denote the treatment indicator for each subject, with $A \in \mathcal{A}=\{0,1\}$ (e.g., insulin and GLP-1 RA). Let $\mathbf{X}=(X_1,\dots, X_p)^\top$ be a $p\times 1$ vector of subject characteristics recorded prior to assigning treatment, and $Y$ be the observed outcome of interest (e.g., blood glucose). The observed data
are $\{(Y_i, A_i, \mathbf{X}^{\top}_i)\}_{i=1}^n$, assumed to be independent and identically distributed across $i$. For brevity, we suppress the subject index $i$ in the following text when no confusion exists.

Under the counterfactual framework for causal inference \cite{robins1986}, we denote $Y(1)$ as the counterfactual outcome that would be observed for a subject if, possibly contrary to fact, treatment $A=1$ were assigned and $Y(0)$ as the counterfactual outcome if treatment $A=0$ were assigned. Then the ATE is
\begin{equation*}
\tau = E\{Y(1)-Y(0)\}.
\end{equation*}
Suppose the marginal density of $\mathbf{X}$, denoted as $f(\mathbf{X})$, exists. Instead of focusing on the entire population as in $\tau$, one may be interested in a specific target population, which can be represented by $f(\mathbf{X})h(\mathbf{X})/E \{ h(\mathbf{X})\}$, where $h(\cdot)$ is a prespecified function of $\mathbf{X}$ that determines the distribution of the target population. In other words, $f(\mathbf{X})h(\mathbf{X})/E \{ h(\mathbf{X})\}$ is the marginal density of $\mathbf{X}$ in the target population, and the function $h(\mathbf{X})$ is to ensure external validity. Then the WATE can be expressed as
\begin{equation}
\label{eq:WATE}
\tau_h = \frac{E\left[h(\mathbf{X})\{Y(1)-Y(0)\}\right]}{E\{h(\mathbf{X})\}}.
\end{equation}
The ATE is a special case of $\tau_h$ with $h(\mathbf{X})=1$.

Note that $\tau_h$ is defined by counterfactual effects; for a given subject, $Y(1)$ and $Y(0)$ cannot both be observed. To connect the counterfactual outcomes with the observed data, we make the three standard assumptions \cite{murphy2001, robins2009}:
\begin{enumerate}
	\item Consistency: $Y=AY(1)+(1-A)Y(0)$, meaning that the observed outcome is the same as the counterfactual outcome under the treatment a patient is actually given, implying no interference between subjects.
	\item Strong ignorability: $A \indep \{Y(1),Y(0)\} | \mathbf{X}$, meaning that the treatment assignment is independent of the counterfactual outcomes, conditional on pre-treatment variables implying that there is no unmeasured confounding. 
	\item Positivity: $\Pr\{\pi(\mathbf{X})\in (0,1)\}=1$ with $\pi(\mathbf{X})=\Pr(A=1|\mathbf{X})$.
\end{enumerate}

$\pi(\mathbf{X})$ is commonly referred to as the propensity score which is often unknown, and thus it is customary to posit a model $\pi(\mathbf{X},\alpha)$, such as logistic regression, to estimate it. We denote $\hat{\alpha}$ as a consistent estimator for $\alpha$ based on observed data, using, for example, maximum likelihood estimation. Similarly, the conditional mean $m_A(\mathbf{X})=E(Y|A,\mathbf{X})$ can be modeled by $m_A(\mathbf{X},\beta)$, and the function $h(\mathbf{X})$ for the target population can be modeled by $h(\mathbf{X},\gamma)$. We denote $\hat{\beta}$ and $\hat{\gamma}$ as consistent estimators for $\beta$ and $\gamma$, respectively, using observed data.

\subsection{Estimation of the WATE with correctly specified $h(\mathbf{X},\gamma)$} \label{h_known}
In this section, we estimate the WATE given that the target function $h(\mathbf{X},\gamma)$ is correctly specified. This also includes the case where $h(\mathbf{X})$ can be fully specified without unknown parameters, for example, $h(\mathbf{X})=I(Age>50)$  with $I(\cdot) $ as a binary indicator function. We derive three estimators of the WATE based on outcome regression, IPW and AIPW.

First of all, it is easy to see that $E\{h(\mathbf{X})Y(A)\}=E[E\{h(\mathbf{X})Y(A)|\mathbf{X}\}]=E\{h(\mathbf{X})m_A(
\mathbf{X})\}$. Hence, an estimator of $\tau_h$ is
\begin{equation}
\label{tau_reg}
\hat{\tau}_h^R = \frac{\sum_ih(\mathbf{X}_i,\hat{\gamma})\{m_1(\mathbf{X}_i,\hat{\beta})-m_0(\mathbf{X}_i,\hat{\beta})\}}{\sum_ih(\mathbf{X}_i,\hat{\gamma})},
\end{equation}
where the superscript $R$ represents outcome regression which can be parametric or nonparametric.

Alternatively, we can estimate $\tau_h$ following the idea of IPW. Defining weights
\begin{equation}
\label{wts}
\left\{
\begin{array}{ll}
w_1(\mathbf{X})=\frac{h(\mathbf{X})}{\pi(\mathbf{X})}, \text{ for } A=1\\
w_0(\mathbf{X})=\frac{h(\mathbf{X})}{1-\pi(\mathbf{X})}, \text{ for } A=0,\\
\end{array}\right.
\end{equation}
we have
\begin{eqnarray*}
	E\{AYw_1(\mathbf{X})\} &=& E\{Y(1)h(\mathbf{X})\}, \\
	E\{(1-A)Yw_0(\mathbf{X})\} &=& E\{Y(0)h(\mathbf{X})\},
\end{eqnarray*}
and
\begin{equation*}
E\{Aw_1(\mathbf{X})\} = E\{(1-A)w_0(\mathbf{X})\} = E\{h(\mathbf{X})\}.
\end{equation*}
Therefore, by equation \eqref{eq:WATE}, an IPW estimator of $\tau_h$ is
\begin{equation}
\label{tau_ipw1}
\hat{\tau}_h^I = \frac{\sum_iA_iY_i w_1(\mathbf{X}_i,\hat{\alpha},\hat{\gamma})}{\sum_iA_i w_1(\mathbf{X}_i,\hat{\alpha},\hat{\gamma})}-\frac{\sum_i(1-A_i)Y_iw_0(\mathbf{X}_i,\hat{\alpha},\hat{\gamma})}{\sum_i(1-A_i)w_0(\mathbf{X}_i,\hat{\alpha},\hat{\gamma})},
\end{equation}
where $w_A(\mathbf{X}_i,\hat{\alpha},\hat{\gamma})$ is a function of $h(\mathbf{X}_i, \hat{\gamma})$ and $\pi(\mathbf{X}_i,\hat{\alpha})$ as in \eqref{wts}.
Equivalently, we can estimate $\tau_h$ as
\begin{equation}
\label{tau_ipw2}
\hat{\tau}_h^I = \frac{\sum_i h(\mathbf{X}_i,\hat{\gamma})\left\{ \frac{A_iY_i}{\pi(\mathbf{X}_i,\hat{\alpha})} -  \frac{(1-A_i)Y_i}{1-\pi(\mathbf{X}_i,\hat{\alpha})} \right\}}{\sum_ih(\mathbf{X}_i,\hat{\gamma})}.
\end{equation}
$\hat{\tau}_h^I$ in \eqref{tau_ipw1} and \eqref{tau_ipw2} are both consistent but \eqref{tau_ipw1} uses normalized weights and is more stable given a small sample size. We use form \eqref{tau_ipw1} in our simulations.

Following the idea of AIPW intended for improved efficiency and robustness, when $h(\mathbf{X},\gamma)$ is correctly specified, we further derive
\begin{eqnarray*}
	&& E\left\{ \frac{AYh(\mathbf{X})}{\pi(\mathbf{X},\alpha)} - \frac{A-\pi(\mathbf{X},\alpha)}{\pi(\mathbf{X},\alpha)}m_1(\mathbf{X},\beta)h(\mathbf{X}) \right\}\\
	& = & E\left\{Y(1)h(\mathbf{X})\right\} + E\left[\frac{A-\pi(\mathbf{X},\alpha)}{\pi(\mathbf{X},\alpha)}\{Y(1)-m_1(\mathbf{X},\beta)\}h(\mathbf{X}) \right]\\
	& = & E\left\{Y(1)h(\mathbf{X})\right\},
\end{eqnarray*}
where the last equation holds if either the propensity score model $\pi(\mathbf{X},\alpha)$ or the conditional mean model $m_1(\mathbf{X},\beta)$ is correctly specified. Similarly, we derive
\begin{eqnarray*}
	&& E\left\{ \frac{(1-A)Yh(\mathbf{X})}{1-\pi(\mathbf{X},\alpha)} + \frac{A-\pi(\mathbf{X},\alpha)}{1-\pi(\mathbf{X},\alpha)}m_0(\mathbf{X},\beta)h(\mathbf{X}) \right\}\\
	& = & E\left\{Y(0)h(\mathbf{X})\right\} + E\left[\frac{A-\pi(\mathbf{X},\alpha)}{1-\pi(\mathbf{X},\alpha)}\{m_0(\mathbf{X},\beta)-Y(0)\}h(\mathbf{X}) \right]\\
	& = & E\{Y(0)h(\mathbf{X})\},
\end{eqnarray*}
where the last equation holds if either the propensity score model $\pi(\mathbf{X},\alpha)$ or the conditional mean model $m_0(\mathbf{X},\beta)$ is correctly specified. 

Therefore, when $h(\mathbf{X},\gamma)$ is correctly specified, we propose an AIPW estimator for $\tau_h$, as
\begin{dmath}
	\label{tau_aipw}
	\hat{\tau}_h^A= \frac{1}{\sum_ih(\mathbf{X}_i,\hat{\gamma})}\sum_ih(\mathbf{X}_i,\hat{\gamma})\left[\left\{ \frac{A_iY_i}{\pi(\mathbf{X}_i,\hat{\alpha})} - \frac{A_i-\pi(\mathbf{X}_i,\hat{\alpha})}{\pi(\mathbf{X}_i,\hat{\alpha})}m_1(\mathbf{X}_i,\hat{\beta}) \right\}-\left\{ \frac{(1-A_i)Y_i}{1-\pi(\mathbf{X}_i,\hat{\alpha})} + \frac{A_i-\pi(\mathbf{X}_i,\hat{\alpha})}{1-\pi(\mathbf{X}_i,\hat{\alpha})}m_0(\mathbf{X}_i,\hat{\beta}) \right\}\right].
\end{dmath}

\begin{theorem} \label{theorem:PropertiesOfTau_h_A}
	Given the assumptions of consistency, strong ignorability and positivity, and a correctly specified $h(\mathbf{X},\gamma)$, $\hat{\tau}_h^A$ in \eqref{tau_aipw} has the following properties:
	\begin{enumerate}
		\item \textbf{Double Robustness}: $\hat{\tau}_h^A$ is a consistent estimator of the $\tau_h$ if either the propensity score model $\pi(\mathbf{X},\alpha)$ or the conditional mean model $m_A(\mathbf{X},\beta)$ is correctly specified.
		\item \textbf{Improved Efficiency}: when $\pi(\mathbf{X},\alpha)$ is correctly specified, if $m_A(\mathbf{X},\beta)$ is also correctly specified, $\hat{\tau}_h^A$ achieves better efficiency than $\hat{\tau}_h^I$.
	\end{enumerate}
\end{theorem}
The first property is obvious by the construction of $\hat{\tau}_h^A$. To see the second property, given correctly specified $h(\mathbf{X},\gamma)$ and $\pi(\mathbf{X},\alpha)$, using the total variance formula $var(\cdot)=E[var\{\cdot|\mathbf{X},Y(1),Y(0)\}] + var[E\{\cdot|\mathbf{X},Y(1),Y(0)\}]$, we have
\begin{equation*}
\resizebox{.95 \textwidth}{!}
{$
	var\left[\left\{ \frac{AYh(\mathbf{X})}{\pi(\mathbf{X})} - \frac{A-\pi(\mathbf{X})}{\pi(\mathbf{X})}m_1(\mathbf{X},\beta)h(\mathbf{X}) \right\}-\left\{ \frac{(1-A)Yh(\mathbf{X})}{1-\pi(\mathbf{X})} + \frac{A-\pi(\mathbf{X})}{1-\pi(\mathbf{X})}m_0(\mathbf{X},\beta)h(\mathbf{X}) \right\}\right]
	$}
\end{equation*}
equal to
\begin{dmath*}
	E\left[ \pi(\mathbf{X})\{1-\pi(\mathbf{X})\}h^2(\mathbf{X})\left\{ \frac{Y(1)-m_1(\mathbf{X},\beta)}{\pi(\mathbf{X})} + \frac{Y(0)-m_0(\mathbf{X},\beta)}{1-\pi(\mathbf{X})} \right\}^2 \right] + var[\{Y(1)-Y(0)\}h(\mathbf{X})].
\end{dmath*}
It is easy to see that the above variance achieves minimal when the conditional mean model $m_A(\mathbf{X},\beta)$ is correctly specified. Since $\hat{\tau}_h^I$ is a special case of $\hat{\tau}_h^A$ with $m_A(\mathbf{X},\beta)=0$, we have the second property. 

\textbf{Remark}. With $h(\mathbf{X})=\pi(\mathbf{X})\{1-\pi(\mathbf{X})\}$, the weights $w_1(\mathbf{X})=1-\pi(\mathbf{X})$ for $A=1$ and $w_0(\mathbf{X})=\pi(\mathbf{X})$ for $A=0$ are called overlap weights in \cite{li2014}, which lead to $\tau_h$ with the smallest asymptotic variance among all WATEs if $\pi(\mathbf{X},\alpha)$ is correctly specified \cite{crump2006, li2014}. By Theorem \ref{theorem:PropertiesOfTau_h_A} Property 2, we could achieve the smallest asymptotic variance by using $\hat{\tau}_h^A$ given correctly specified $\pi(\mathbf{X},\alpha)$ and $m_A(\mathbf{X},\beta)$. Moreover, the overlap weights can balance the small-sample means of all covariates involved in the propensity score model when the propensity scores are estimated by maximum likelihood under a logistic model \cite{li2014}. In addition, the overlap weights are stable in the presence of extreme propensity scores.

\subsection{Estimation of the WATE with correctly specified $h(\mathbf{X},\gamma)$} \label{h_unknown}
The WATE $\tau_h$ defines a general class of estimands by using $h$ functions for various target populations. When $h(\mathbf{X},\gamma)$ can be correctly specified, we have the doubly robust estimator $\hat{\tau}_h^A$ for $\tau_h$ as shown in Section \ref{h_known}. In some scenarios, we know the form of $h(\mathbf{X})$ but could not guarantee the correct specification of $h(\mathbf{X},\gamma)$. For example, in an observational study, if $h(\mathbf{X})$ is a function of $\pi(\mathbf{X})$,  $h(\mathbf{X},\gamma)$ is subject to the misspecification of $\pi(\mathbf{X},\alpha)$. Then the results in Section \ref{h_known} may not hold.

Two important cases are the ATT and ATC. Given $h(\mathbf{X})=\pi(\mathbf{X})$, $\tau_h$ is equal to the ATT, i.e.,
\begin{equation*} \label{eq:ATT}
\tau_{\pi} = \frac{E\left[\pi(\mathbf{X})\{Y(1)-Y(0)\}\right]}{E\{\pi(\mathbf{X})\}} = E\{Y(1)-Y(0)|A=1\}.
\end{equation*} 
Given $h(\mathbf{X})=1-\pi(\mathbf{X})$, $\tau_h$ is equal to the ATC, i.e.,
\begin{equation*} \label{eq:ATC}
\tau_{1-\pi} =\frac{E\left[\{1-\pi(\mathbf{X})\}\{Y(1)-Y(0)\}\right]}{E\{1-\pi(\mathbf{X})\}} = E\{Y(1)-Y(0)|A=0\}.
\end{equation*}

With $h(\mathbf{X})$ depending on $\pi(\mathbf{X})$, the estimator $\hat{\tau}_h^A$ for the ATT or the ATC is no longer doubly robust since a misspecified propensity model may lead to a wrong target population. However, it is straightforward to figure out the regression-based and IPW estimators separately for both the ATT and the ATC. Specifically, for the ATT, we have
\begin{equation*}
\hat{\tau}_{\pi}^R=\frac{\sum_iA_i\{Y_i - m_0(\mathbf{X}_i, \hat{\beta})\}}{\sum_iA_i}  \text{ and } \hat{\tau}_{\pi}^I=\frac{\sum_i\{A_iY_i-\frac{\pi(\mathbf{X}_i,\hat{\alpha})(1-A_i)}{1-\pi(\mathbf{X}_i,\hat{\alpha})}Y_i \}}{\sum_i\pi(\mathbf{X}_i,\hat{\alpha})}.
\end{equation*}
For the ATC, we have
\begin{equation*}
\hat{\tau}_{1-\pi}^R=\frac{\sum_i(1-A_i)\{m_1(\mathbf{X}_i,\hat{\beta}) - Y_i\}}{\sum_i(1-A_i)}
\text{ and }
\hat{\tau}_{1-\pi}^I=\frac{\sum_i\{\frac{1-\pi(\mathbf{X}_i,\hat{\alpha})}{\pi(\mathbf{X}_i,\hat{\alpha})}A_iY_i-(1-A_i)Y_i \}}{\sum_i\{1-\pi(\mathbf{X}_i,\hat{\alpha})\}}.
\end{equation*}

By combining the regression-based and IPW estimators, we find the following doubly robust estimators for the ATT and the ATC.
\begin{prop} \label{prop:att_atc}
	Given the assumptions of consistency, strong ignorability and positivity, $\hat{\tau}_{\pi}^{DB}$ with the form
	\begin{equation}
	\hat{\tau}_{\pi}^{DR}=\frac{\sum_i\left[A_iY_i -\left\{ \frac{\pi(\mathbf{X}_i,\hat{\alpha})(1-A_i)}{1-\pi(\mathbf{X}_i,\hat{\alpha})}Y_i + \frac{A_i-\pi(\mathbf{X}_i,\hat{\alpha})}{1-\pi(\mathbf{X}_i,\hat{\alpha})}m_0(\mathbf{X}_i,\hat{\beta})\right\} \right] }{\sum_iA_i}
	\end{equation}
	is a consistent estimator of the ATT if either the propensity score model $\pi(\mathbf{X},\alpha)$ or the conditional mean model $m_0(\mathbf{X},\beta)$ is correctly specified,
	and $\hat{\tau}_{1-\pi}^{DB}$ with the form
	\begin{equation}
	\hat{\tau}_{1-\pi}^{DR}=\frac{\sum_i\left[\left\{\frac{1-\pi(\mathbf{X}_i,\hat{\alpha})}{\pi(\mathbf{X}_i,\hat{\alpha})}A_iY_i - \frac{A_i-\pi(\mathbf{X}_i,\hat{\alpha})}{\pi(\mathbf{X}_i,\hat{\alpha})}m_1(\mathbf{X}_i,\hat{\beta}) \right\} - (1-A_i)Y_i \right] }{\sum_i(1-A_i)}
	\end{equation}
	is a consistent estimator of the ATC if either the propensity score model $\pi(\mathbf{X},\alpha)$ or the conditional mean model $m_1(\mathbf{X},\beta)$ is correctly specified.
\end{prop}

Moreover, we consider a general form $h(\mathbf{X})=a+b\pi(\mathbf{X})$, where $a$ and $b$ are known constants with $a^2+b^2>0$. Then the ATT and the ATC are special cases with $a=0, b=1$ and $a=1,b=-1$. We have the following theorem for a doubly robust estimator of $\tau_h$. The proof is in the Appendix, which also implies Proposition \ref{prop:att_atc}.

\begin{theorem} \label{theorem:DRhfpi}
	Given the assumptions of consistency, strong ignorability and positivity, and $h(\mathbf{X})=a+b\pi(\mathbf{X})$, $\hat{\tau}^{DR}_h$ with the form
	\begin{dmath*}
		\hat{\tau}^{DR}_h =\frac{1}{\sum_i(a+bA_i)}\sum_i \left((a+bA_i)\{m_1(\mathbf{X}_i,\hat{\beta})-m_0(\mathbf{X}_i,\hat{\beta})\} + \{a+b \pi(\mathbf{X}_i,\hat{\alpha})\}\left[\frac{A_i}{\pi(\mathbf{X}_i,\hat{\alpha})}\{Y_i-m_1(\mathbf{X}_i,\hat{\beta})\} -\frac{1-A_i}{1-\pi(\mathbf{X}_i,\hat{\alpha})}\{Y_i-m_0(\mathbf{X}_i,\hat{\beta})\}\right] \right)
	\end{dmath*}
	is a consistent estimator of $\tau_h$ if either the propensity score model $\pi(\mathbf{X},\alpha)$ or the conditional mean model $m_A(\mathbf{X},\beta)$ is correctly specified.
\end{theorem}

For the ATO, denoting the probability limit of the propensity score under the misspecified model as $\pi^{*}(\mathbf{X})$ and assuming a correctly specified conditional mean model, the ATO estimator is consistent for 
\begin{equation}
\label{misATO}
\frac{E\left[h^{*}(\mathbf{X})\{Y(1)-Y(0)\}\right]}{E\{h^{*}(\mathbf{X})\}},
\end{equation}
with $h^{*}(\mathbf{X})=\pi^{*}(\mathbf{X})\{1- \pi^{*}(\mathbf{X})\}$. In other words, it retains a causal interpretation with a misspecified propensity score model and a correctly specified conditional mean model. Given an intercept-only model for the propensity score, equivalent to using sample proportions, the estimand \eqref{misATO} is reduced to the ATE. By using a model that explains only part of the total variance in the treatment assignment, we expect the estimand \eqref{misATO} to yield an estimate somewhere between the ATE and ATO estimates when these two are estimated under correct models for both $\pi$ and $m$. Therefore, as a trade-off in certain scenarios, one may favor the estimand \eqref{misATO} due to its stableness against extreme weights and asymptotic efficiency.

\section{Simulation studies}

We conduct simulation studies to evaluate the performance of the proposed estimators under four possible combinations of correctly and incorrectly specified propensity score model and conditional mean model. We consider estimands of the ATE, ATT, ATC and ATO, with $\hat{\tau}^{A}_h$ for the ATE and ATO and $\hat{\tau}^{DR}_h$ for the ATT and ATC. For comparison, we also apply the corresponding regression-based and IPW estimators for each estimand, with the criteria being average bias and root mean square error (RMSE) over 1000 Monte Carlo replications. We consider sample sizes of $200$ and $1000$.

We simulate five covariates $\mathbf{X}=(X_1,\dots,X_5)^\top$ following $N(0,I_5)$ where $I_5$ is a $5\times 5$ identity matrix. Treatment $A$ is generated from $Binomial\{\pi(\mathbf{X})\}$ with $logit\{\pi(\mathbf{X})\} = 0.5+X_1-0.5X_2^2+0.5X_3X_5$ and $logit(\pi)=\log\{\pi/(1-\pi)\}$. Outcome $Y$ is generated from two difference models:
\begin{enumerate}
	\item $Y = 1+X_2^2+X_3+A\exp(X_1+0.5X_3X_5) + \epsilon$,
	\item $Y = 1+X_2^2+X_3+A(X_1+0.5X_3X_5) + \epsilon$,
\end{enumerate}
with $\epsilon \sim N(0,1)$. Outcome model 1 indicates an overall advantage of treatment $A=1$ over treatment $A=0$ while outcome model 2 indicates no overall difference. The true values for the ATE, ATT, ATC and ATO are $1.90$, $2.75$, $1.04$ and $1.50$, respectively, under outcome model 1, and $0$, $0.46$, $-0.46$ and $0$ under outcome model 2. Note that the ATE and ATO are different under outcome model 1 but the same under outcome model 2, due to the fact that $Y(1)$ under outcome model 1 is skewed to have more large values while $Y(1)$ under outcome model 1 and $Y(0)$ under both models are more symmetric.

The misspecified outcome regression model is $m^{'}(\mathbf{X}) = \gamma_0 + \sum_{k=1}^{5}\gamma_kX_k + A(\eta_0 + \sum_{j=1}^{5}\eta_jX_j)$, which leads to $R^2$ around $0.2$ for outcome model 1 and $0.3$ for outcome model 2 (The correctly specified model has $R^2$ around $0.9$ for outcome model 1 and $0.8$ for outcome model 2). The misspecified propensity model is $logit\{\pi^{'}(\mathbf{X}) \} = \zeta_0 + \sum_{k=1}^{5}\zeta_kX_k$, leading to a correlation around $0.75$ between the linear predictors $logit\{\hat{\pi}^{'}(\mathbf{X})\}$ and the truth $logit\{\pi(\mathbf{X})\}$.

Table \ref{simu1} summarizes the performances of all estimators considered under outcome model 1. When both propensity score and conditional mean models are correctly specified, all estimators have minimal bias and RMSE, with $\hat{\tau}_h^R$ and the proposed estimators ($\hat{\tau}^{A}_h$ for the ATE and ATO and $\hat{\tau}^{DR}_h$ for the ATT and ATC) both more efficient than $\hat{\tau}_h^I$. When only $\pi(\mathbf{X},\alpha)$ is correctly specified, the proposed estimators are consistent despite moderate biases for the ATE and ATC under sample size $200$. When only $m_A(\mathbf{X},\beta)$ is correctly specified, the proposed estimators yield unbiased estimates for the ATE, ATT and ATC with RMSE similar to $\hat{\tau}_h^R$. However, for the ATO, there is moderate bias because its target function $h(\mathbf{X},\gamma)=\pi(\mathbf{X},\alpha)\{1-\pi(\mathbf{X},\alpha)\}$ is misspecified. Nonetheless, we do observe the improved efficiency of the ATO over the ATE. With at least one model correctly specified, our proposed estimators have either similar or better performance compared to $\hat{\tau}_h^I$. In addition, increasing the sample size reduces the bias and RMSE in general, with $\hat{\tau}_h^I$ and the proposed estimators with only $\pi(\mathbf{X},\alpha)$ correct having the most improvement. 

Results under outcome model 2 are presented in Table \ref{simu2}. Similar to Table \ref{simu1}, Table \ref{simu2} corroborates the double robustness of the proposed estimators for the ATE, ATT and ATC. When $\pi(\mathbf{X},\alpha)$ is correctly specified, our proposed estimators with a misspecified but informative $m_A(\mathbf{X},\beta)$ have similar RMSE as $\hat{\tau}_h^I$. When $m_A(\mathbf{X},\beta)$ is correctly specified, our proposed estimators have similar RMSE as $\hat{\tau}_h^R$, which is better than $\hat{\tau}_h^I$, regardless of the specification of $\pi(\mathbf{X},\alpha)$. The major difference between Table \ref{simu1} and Table \ref{simu2} lies in the ATO estimates given misspecified $\pi(\mathbf{X},\alpha)$ and correctly specified $m_A(\mathbf{X},\beta)$. $\hat{\tau}^{A}_h$ is not doubly robust for the ATO since its target function $h(\mathbf{X},\gamma)$ depends on $\pi(\mathbf{X},\alpha)$. As mentioned in Section \ref{h_unknown}, when the misspecified $\pi(\mathbf{X},\alpha)$ explains partially the variation, we expect the ATO estimate falls somewhere between the ATE and ATO estimates given correctly specified models for both $\pi$ and $m$. This likely explains why we observe positive bias when the ATE is greater than the ATO under outcome model 1, and a consistent estimate when the ATE and ATO are the same under outcome model 2. Under both outcome models, $\hat{\tau}^{A}_h$ for the ATO performs similarly to, if not better than, $\hat{\tau}^{I}_h$, with a correctly specified $m_A(\mathbf{X},\beta)$ leading to improved performance of $\hat{\tau}^{A}_h$ over $\hat{\tau}^{I}_h$ given the same $\pi(\mathbf{X},\alpha)$. In summary, despite not having double robustness for the ATO, $\hat{\tau}^{A}_h$ shows to be a better choice than $\hat{\tau}^{I}_h$.

\section{Application}

In this section we apply our proposed estimators to the CPRD data featuring patients with T2D from the UK. We include patients of age $\ge 21$ who were active (i.e., registered at a CPRD practice) during the study period from January 1, 2011 to December 31, 2014. Treatment of interest is the GLP-1 RA therapy, which is a relatively new first-line injectable anti-diabetic treatment for T2D patients, compared to the standard of care of insulin therapy. GLP-1 RA therapy usually works as a once daily or weekly injectable, as opposed to multiple daily injections for common insulin therapy, and has been shown with lower risk of hypoglycemia and better weight control \cite{wang2011, abdul2013}. Outcome is the reduction of hemoglobin A1c (HbA1c, \%) six months after initiation of the treatment.

We have 436 patients on insulin and 192 on GLP-1 RA. Patients on insulin were overall older than patients on GLP-1 RA (mean: 61.7 years vs. 54.5 years) with higher baseline HbA1c (mean: 10.4\% vs. 9.5\%). Another major difference was that patients on GLP-1 RA had much higher BMI (mean: 39.1 vs. 29.2). This is consistent with previous studies as physicians would more likely prescribe GLP-1 RA for patients with weight issues due to its significantly better weight control \cite{wang2011, abdul2013}. We aim to investigate whether the GLP-1 RA therapy comes at the cost of worse glycemic control than insulin therapy for those target patients by applying the ATT estimand. Another goal is to gain insight into whether switching current insulin users to GLP-1 RA in order to reduce treatment burden is a reasonable treatment recommendation, for which we will apply the ATC estimand. Other baseline covariates include gender, smoking status, lipid levels, comorbidities and use of oral anti-diabetic medications. Missing data is imputed using R package \emph{mice}. We fit a logistic regression model with main effects of all baseline covariates for the propensity scores and a linear regression model with main effects of treatment and baseline covariates as well as the interaction terms of treatment by baseline covariates for the conditional means. To avoid extreme weights, we truncate the estimated propensity scores by the 5th and 95th percentiles.

The WATE estimates with bootstrap standard errors (SEs) are shown in Table \ref{dataapp}. Due to covariate imbalance in the two treatment groups, we observe very different results before and after covariate adjustment. Without weighting, the insulin group has significantly better HbA1c reduction than the GLP-1 group, with an estimated mean difference of 1.06\%. In contrast, the estimates for all estimands are much smaller after adjustment for covariate imbalance, yielding no statistical significance. Our proposed estimators result in smaller estimates than $\hat{\tau}^{R}_h$ and $\hat{\tau}^{I}_h$ for all estimands. The ATE estimate of 0.13\% means that patients on insulin have on average 0.13\% more HbA1c reduction than those on GLP-1 RA. Existing meta-analyses have discovered either no significant difference between the two treatment groups \cite{abdul2013}, or slightly higher HbA1c reduction in the insulin group \cite{wang2011}. An estimate of 0.14\% reported by Wang \emph{et al.}\cite{wang2011} is very close to the ATE estimate by our proposed estimator $\hat{\tau}^{A}_h$, although our SE is much larger likely due to a smaller sample size. In particular, the ATT estimate is the closest to zero, meaning that these younger and more overweight patients with lower baseline HbA1c would have similar glycemic control with either treatment. Therefore, our study supports the prescription of GLP-1 RA as opposed to insulin for these target patients so as to enjoy the better weight control and lower rate of hypoglycemia without sacrificing the glycemic control \cite{wang2011, abdul2013}. The nonsignificant ATC estimate of 0.19\% supports the idea of switching current insulin users to GLP-1 RA for less frequent injection. As expected, the ATO estimates have smaller SEs than all other estimands.

\section{Discussion}

We have proposed estimators for a general class of WATEs defined by the target function $h$, which is flexible in estimating the ATE for a target population while controlling for covariate imbalance in an observational study. By combining the regression-based and IPW estimators, we have proposed new estimators for various WATEs. Given a correctly specified $h(\mathbf{X},\gamma)$, the proposed estimator $\hat{\tau}^{A}_h$ is doubly robust, meaning that it is consistent if either the propensity score model $\pi(\mathbf{X},\alpha)$ or the conditional mean model $m_A(\mathbf{X},\beta)$ is correctly specified. It also enjoys better efficiency than the IPW estimator $\hat{\tau}^{I}_h$ if both $\pi(\mathbf{X},\alpha)$ and $m_A(\mathbf{X},\beta)$ are both correctly specified. Moreover, our simulation studies show that $\hat{\tau}^{A}_h$ has either similar or better efficiency compared to $\hat{\tau}^{I}_h$ if at least one of the two models is correctly specified, and it is also a better choice for the ATO even without double robustness. When $h(\mathbf{X},\gamma)$ is a linear function of $\pi(\mathbf{X},\alpha)$, which is subject to model misspecification, we have proposed a doubly robust estimator $\hat{\tau}^{DR}_h$, which can be used to estimate the ATT and ATC, two common estimands in practice.

The proposed estimators can be further improved in several directions. First, one can incorporate the idea by Han and Wang \cite{han2013} to allow multiple models for both the propensity score and the conditional mean so as to go beyond double robustness. Second, the propensity score estimation can be modified to improve the empirical performance of propensity score weighting methods, for example, the covariate balancing propensity score which models treatment assignment while
optimizing the covariate balance \cite{imai2014}. Third, following the method by Cao \emph{et al.} \cite{cao2009}, one can explore alternative doubly robust estimators that can minimize the asymptotic variance with correctly specified $\pi(\mathbf{X},\alpha)$ but misspecified $m_A(\mathbf{X},\beta)$, and meanwhile apply restriction to drive estimated propensity scores away from zero. Last but not least, instead of focusing on variance reduction, it may be of interest to reduce bias following the bias-reduced doubly robust estimation by Vermeulen and Vansteelandt \cite{vermeulen2015}.

\section*{Appendix}\label{sec:appendix}
\emph{Proof of Theorem \ref{theorem:DRhfpi}.} If the propensity score model $\pi(\mathbf{X},\alpha)$ is correctly specified, we have
\begin{eqnarray*}
	E\left[(a+b A)\{m_1(\mathbf{X},\beta)-m_0(\mathbf{X},\beta)\}\right] &=& E\left[E(a+bA|\mathbf{X})\{m_1(\mathbf{X},\beta)-m_0(\mathbf{X},\beta)\} \right] \\
	&=& E\left[ \{a+b\pi(\mathbf{X})\}  \{m_1(\mathbf{X},\beta)-m_0(\mathbf{X},\beta)\}\right],
\end{eqnarray*}
and
\begin{eqnarray*}
	&&E\left(\{a+b \pi(\mathbf{X})\}\left[\frac{A}{\pi(\mathbf{X})}\{Y-m_1(\mathbf{X},\beta)\}-\frac{1-A}{1-\pi(\mathbf{X})}\{Y-m_0(\mathbf{X},\beta)\}\right]\right)\\
	&=&E \left[ \{a+b \pi(\mathbf{X})\} \left\{\frac{AY}{\pi(\mathbf{X})} -\frac{(1-A)Y}{1-\pi(\mathbf{X})}\right\}\right]  -E\left[ \{a+b\pi(\mathbf{X})\}  \{m_1(\mathbf{X},\beta)-m_0(\mathbf{X},\beta)\}\right],\\
	&=&  E \left[h(\mathbf{X})\{Y(1)-Y(0)\}\right] - E\left[ \{a+b\pi(\mathbf{X})\}  \{m_1(\mathbf{X},\beta)-m_0(\mathbf{X},\beta)\}\right],
\end{eqnarray*}
which implies $ \hat{\tau}^{DR}_h \rightarrow \tau_h$.

If the conditional mean model $m_A(\mathbf{X},\beta)$ is correctly specified, we have
\begin{eqnarray*}
	&&E\left(\{a+b \pi(\mathbf{X},\alpha)\}\left[\frac{A}{\pi(\mathbf{X},\alpha)}\{Y-m_1(\mathbf{X})\}-\frac{1-A}{1-\pi(\mathbf{X},\alpha)}\{Y-m_0(\mathbf{X})\}\right]\right)\\
	&=& E\left(\{a+b \pi(\mathbf{X},\alpha)\}E\left[\frac{A}{\pi(\mathbf{X},\alpha)}\{Y(1)-m_1(\mathbf{X})\}-\frac{1-A}{1-\pi(\mathbf{X},\alpha)}\{Y(0)-m_0(\mathbf{X})\}\middle| \mathbf{X}\right] \right) \\
	&=& E\left(\{a+b \pi(\mathbf{X},\alpha)\}\left[\frac{\pi(\mathbf{X})}{\pi(\mathbf{X},\alpha)}E\{Y(1)-m_1(\mathbf{X})| \mathbf{X}\}-\frac{1-\pi(\mathbf{X})}{1-\pi(\mathbf{X},\alpha)}E\{Y(0)-m_0(\mathbf{X})| \mathbf{X}\}\right] \right) \\
	&=& 0,
\end{eqnarray*}
and 
\begin{eqnarray*}
	E\left[(a+b A)\{m_1(\mathbf{X})-m_0(\mathbf{X})\}\right] &=& E\left[\{a+\pi(\mathbf{X})\}E\{Y(1)-Y(0)|\mathbf{X}\}\right] \\
	&=& E \left[h(\mathbf{X})\{Y(1)-Y(0)\}\right],
\end{eqnarray*}
which implies $ \hat{\tau}^{DR}_h \rightarrow \tau_h$. $\Box$

\newpage
\bibliography{ref}  
\bibliographystyle{biom} 

\newpage
\begin{table}
	\fontsize{10}{12}\selectfont
	\begin{center}
		\caption{Simulation results for WATE $\tau_h$ under outcome model 1, based on $1000$ Monte Carlo replications. $\pi$ is the propensity score model and $m$ is the conditional mean model for the outcome. The marks of $\checkmark$ and $\times$ indicate correct and incorrect model specifications, respectively. Bias is the average difference between the estimate and the truth. RMSE is the square root of the average squared difference between the estimate and the truth.}\label{simu1}
		\resizebox{\textwidth}{!}{\begin{tabularx}{0.85\textwidth}{c *{11}{Y}}
				\toprule
				\multirow{2}{*}{$\pi$} & \multirow{2}{*}{$m$} & \multirow{2}{*}{Method} & \multicolumn{2}{c}{ATE}  & \multicolumn{2}{c}{ATT} & \multicolumn{2}{c}{ATC} & \multicolumn{2}{c}{ATO}\tabularnewline
				\cmidrule(l){4-5}\cmidrule(l){6-7}\cmidrule(l){8-9}\cmidrule(l){10-11}
				& & & Bias & RMSE & Bias & RMSE & Bias & RMSE & Bias & RMSE \tabularnewline
				\midrule
				& & & \multicolumn{8}{c}{$n=200$}\tabularnewline
				\cmidrule(l){4-11}
				\multirow{2}{*}{-} & $\checkmark$ & \multirow{2}{*}{$\hat{\tau}_h^{R}$} & -0.01 & 0.30 & -0.03 & 0.51  & 0.00 & 0.21 & - & - \tabularnewline
				& $\times$ & & -0.93 & 1.02 & -0.81 & 0.98 & -1.06 & 1.23 & - & - \tabularnewline
				\midrule
				$\checkmark$ & \multirow{2}{*}{-} & \multirow{2}{*}{$\hat{\tau}_h^I$} & -0.07 & 0.42 & -0.03 & 0.56  & -0.13 & 0.54 & -0.01 & 0.28 \tabularnewline
				$\times$ &  &  & -0.63 & 0.76 & -0.85 & 1.05  & -0.41 & 0.55 & -0.46 & 0.58\tabularnewline
				\midrule
				$\checkmark$ & $\checkmark$ & \multirow{4}{*}{$\hat{\tau}_h^A$ / $\hat{\tau}_h^{DR}$} & -0.01 & 0.31 & -0.03 & 0.51 & 0.00 & 0.28 & -0.01 & 0.22\tabularnewline
				$\checkmark$ & $\times$ &  & -0.11 & 0.44 & -0.04 & 0.53  & -0.21 & 0.77 & -0.03 & 0.26 \tabularnewline
				$\times$ & $\checkmark$ &  & -0.02 & 0.30 & -0.03 & 0.51  & 0.00 & 0.24 & 0.14 & 0.29\tabularnewline
				$\times$ & $\times$ &  & -0.66 & 0.81 & -0.85 & 1.04  & -0.47 & 0.68 & -0.47 & 0.60\tabularnewline
				\midrule
				& & & \multicolumn{8}{c}{$n=1000$}\tabularnewline
				\cmidrule(l){4-11}
				\multirow{2}{*}{-} & $\checkmark$ & \multirow{2}{*}{$\hat{\tau}_h^{R}$} & 0.00 & 0.16 & 0.00 & 0.30  & 0.00 & 0.09 & - & - \tabularnewline
				& $\times$ & & -0.92 & 0.94 & -0.80 & 0.87 & -1.04 & 1.08 & - & - \tabularnewline
				\midrule
				$\checkmark$ & \multirow{2}{*}{-} & \multirow{2}{*}{$\hat{\tau}_h^I$} & -0.02 & 0.25 & 0.00 & 0.33  & -0.05 & 0.34 & 0.00 & 0.12 \tabularnewline
				$\times$ & & & -0.62 & 0.66 & -0.84 & 0.90  & -0.41 & 0.44 & -0.45 & 0.49\tabularnewline
				\midrule
				$\checkmark$ & $\checkmark$ & \multirow{4}{*}{$\hat{\tau}_h^A$ / $\hat{\tau}_h^{DR}$} & 0.00 & 0.17 & 0.00 & 0.30 & 0.00 & 0.13 & 0.00 & 0.09\tabularnewline
				$\checkmark$ & $\times$ & & -0.03 & 0.27 & 0.00 & 0.30  & -0.06 & 0.47 & -0.01 & 0.10\tabularnewline
				$\times$ & $\checkmark$ & & 0.00 & 0.16 & 0.01 & 0.30  & 0.00 & 0.10 & 0.16 & 0.21\tabularnewline
				$\times$ & $\times$ & & -0.64 & 0.68 & -0.84 & 0.90  & -0.44 & 0.48 & -0.46 & 0.49\tabularnewline
				\bottomrule
			\end{tabularx}}
		\end{center}
	\end{table}

\begin{table}
	\fontsize{10}{12}\selectfont
	\begin{center}
		\caption{Simulation results for WATE $\tau_h$ under outcome model 2, based on $1000$ Monte Carlo replications. $\pi$ is the propensity score model and $m$ is the conditional mean model for the outcome. The marks of $\checkmark$ and $\times$ indicate correct and incorrect model specifications, respectively. Bias is the average difference between the estimate and the truth. RMSE is the square root of the average squared difference between the estimate and the truth.}\label{simu2}
		\resizebox{\textwidth}{!}{\begin{tabularx}{0.85\textwidth}{c *{11}{Y}}
				\toprule
				\multirow{2}{*}{$\pi$} & \multirow{2}{*}{$m$} & \multirow{2}{*}{Method} & \multicolumn{2}{c}{ATE}  & \multicolumn{2}{c}{ATT} & \multicolumn{2}{c}{ATC} & \multicolumn{2}{c}{ATO}\tabularnewline
				\cmidrule(l){4-5}\cmidrule(l){6-7}\cmidrule(l){8-9}\cmidrule(l){10-11}
				& & & Bias & RMSE & Bias & RMSE & Bias & RMSE & Bias & RMSE \tabularnewline
				\midrule
				& & & \multicolumn{8}{c}{$n=200$}\tabularnewline
				\cmidrule(l){4-11}
				\multirow{2}{*}{-} & $\checkmark$ & \multirow{2}{*}{$\hat{\tau}_h^{R}$} & 0.00 & 0.18 & -0.01 & 0.20  & 0.00 & 0.21 & - & - \tabularnewline
				& $\times$ & & -0.65 & 0.71 & -0.78 & 0.85 & -0.51 & 0.58 & - & - \tabularnewline
				\midrule
				$\checkmark$ & \multirow{2}{*}{-} & \multirow{2}{*}{$\hat{\tau}_h^I$} & -0.05 & 0.36 & 0.00 & 0.32  & -0.11 & 0.55 & -0.01 & 0.24 \tabularnewline
				$\times$ & & & -0.65 & 0.73 & -0.83 & 0.93  & -0.48 & 0.59 & -0.63 & 0.69\tabularnewline
				\midrule
				$\checkmark$ & $\checkmark$ & \multirow{4}{*}{$\hat{\tau}_h^A$ / $\hat{\tau}_h^{DR}$} & 0.00 & 0.20 & 0.00 & 0.23 & 0.00 & 0.25 & 0.00 & 0.18\tabularnewline
				$\checkmark$ & $\times$ & & -0.09 & 0.33 & -0.02 & 0.30 & -0.15 & 0.55 & -0.03 & 0.19 \tabularnewline
				$\times$ & $\checkmark$ & & 0.00 & 0.19 & -0.01 & 0.21 & 0.00 & 0.22 & 0.01 & 0.18\tabularnewline
				$\times$ & $\times$ & & -0.66 & 0.73 & -0.83 & 0.92  & -0.49 & 0.58 & -0.63 & 0.69\tabularnewline
				\midrule
				& & & \multicolumn{8}{c}{$n=1000$}\tabularnewline
				\cmidrule(l){4-11}
				\multirow{2}{*}{-} & $\checkmark$ & \multirow{2}{*}{$\hat{\tau}_h^{R}$} & 0.00 & 0.08 & 0.00 & 0.09  & 0.00 & 0.09 & - & - \tabularnewline
				& $\times$ & & -0.66 & 0.67 & -0.81 & 0.82 & -0.50 & 0.52 & - & - \tabularnewline
				\midrule
				$\checkmark$ & \multirow{2}{*}{-} & \multirow{2}{*}{$\hat{\tau}_h^I$} & -0.02 & 0.22 & -0.01 & 0.15  & -0.04 & 0.36 & -0.01 & 0.08 \tabularnewline
				$\times$ & & & -0.67 & 0.68 & -0.84 & 0.86  & -0.49 & 0.51 & -0.64 & 0.65\tabularnewline
				\midrule
				$\checkmark$ & $\checkmark$ & \multirow{4}{*}{$\hat{\tau}_h^A$ / $\hat{\tau}_h^{DR}$} & 0.00 & 0.09 & -0.01 & 0.10 & 0.00 & 0.12 & 0.00 & 0.08\tabularnewline
				$\checkmark$ & $\times$ & & -0.03 & 0.20 & -0.01 & 0.12  & -0.04 & 0.38 & -0.01 & 0.08\tabularnewline
				$\times$ & $\checkmark$ & & 0.00 & 0.08 & 0.00 & 0.09  & 0.00 & 0.09 & 0.02 & 0.08\tabularnewline
				$\times$ & $\times$ & & -0.67 & 0.68 & -0.84 & 0.86  & -0.49 & 0.51 & -0.64 & 0.65\tabularnewline
				\bottomrule
			\end{tabularx}}
		\end{center}
	\end{table}

\begin{table}
	\fontsize{10}{12}\selectfont
	\begin{center}
		\caption{Weighted mean (SE) differences in HbA1c (\%) reduction after six months of GLP-1 RA vs. insulin therapies among patients with type 2 diabetes, with positive numbers favoring insulin. The ATT is for the patient population on GLP-1 RA while the ATC is for the patient population on insulin.}\label{dataapp}
		\resizebox{\textwidth}{!}{\begin{tabularx}{0.85\textwidth}{c *{5}{Y}}
				\toprule
				Method & ATE & ATT & ATC & ATO \tabularnewline
				\midrule
				Unweighted & 1.06 (0.20) & - & - & - \tabularnewline
				$\hat{\tau}_h^{R}$ & 0.34 (0.36) & 0.25 (0.36) & 0.38 (0.46) & - \tabularnewline
				$\hat{\tau}_h^I$ & 0.44 (0.45) & 0.12 (0.57) & 0.54 (0.69) & 0.37 (0.29) \tabularnewline
				$\hat{\tau}_h^A$ / $\hat{\tau}_h^{DR}$ & 0.13 (0.43) & -0.02 (0.61) & 0.19 (0.53) & 0.27 (0.31) \tabularnewline
				\bottomrule
			\end{tabularx}}
		\end{center}
	\end{table}

\end{document}